\documentclass[aps,twocolumn,nofootinbib]{revtex4}
\usepackage[latin1]{inputenc}
\usepackage{epsfig}
\newcommand{\beq}{\begin{equation}}
\newcommand{\eeq}{\end{equation}}
\newcommand{\la}{\langle}
\newcommand{\ra}{\rangle}

\begin{document}

\title{Derivation of Fokker-Planck equation and its
entropy production}

\author{Tânia Tomé and Mário J. de Oliveira}

\affiliation{Universidade de São Paulo,
Instituto de Física,
Rua do Matão, 1371, 05508-090
São Paulo, SP, Brasil}

\begin{abstract}

We derive the Fokker-Planck equation and its entropy
production from the master equation. To this end we
consider a discrete space of states where the master
equation is defined and appropriate transition rates
are introduced. A transition rate has two parts,
one of which is the ordinary term proportional to the
inverse of the spacing and the other is inversely
proportional to the square of the spacing corresponding
to the variance of the independent stochastic variable.
In the continuous space limit, we obtain the Fokker-Planck
as well as the rate of entropy production and the
flux of entropy. We also consider the case where the
space of states is the phase space and discuss the
equations appropriate for systems in contact with
several heat reservoirs and for isolated systems.

\end{abstract}

\maketitle

\section{Introdução}

The Fokker-Planck equation
\cite{kampen1981,gardiner1983,risken1984,tome2015L}
governs the time evolution
of the probability distribution defined on a continuous
space of states and is recognized as a Markovian stochastic
dynamics in continous time \cite{kolmogorov1931}.
The equation was originally derived by Fokker
\cite{fokker1914} and by Planck \cite{planck1917}
as a generalization of the equation introduced by
Einstein \cite{einstein1905} to describe the
diffusion of Brownian particles.

The reasoning used by Einstein to reach the equation was
as follows. Let us consider the stochastic motion of a
particle along a straight line. At each time interval
$\tau$ the particle jumps a certain distance to the right
or to the left with the same probability. Denoting by
$x$ and $x'$ the positions of the particle at time $t$
and $t+\tau$, then $\zeta=x'-x$ can be understood as
a randon variable with zero mean and nonzero  
variance $\overline{\zeta^2}$. Denoting $\rho$ and
$\rho'$ the probability distribution of the position
of the particles at time $t$ and $t+\tau$ then
it follows that they are related by
\beq
\rho'-\rho =  \frac{\overline{\zeta^2}}2
\frac{\partial^2\rho}{\partial x^2},
\eeq
Dividing this equation by $\tau$ and taking the 
continuous time limit, it becomes
\beq
\frac{\partial\rho}{\partial t}
= D \frac{\partial^2 \rho}{\partial x^2},
\label{10}
\eeq
where $D$ is defined by
\beq
D = \frac1{2\tau} \overline{\zeta^2}.
\label{11}
\eeq
For the equation (\ref{10}) to make sense, it is
necessary that $D$ be finite from which follows that
the variance $\overline{\zeta^2}$ vanishes with
$\tau$ when $\tau$ decreases without limit, a result
that is implicit in Einstein reasoning.

From this reasoning it also follows the 
well known Einstein formula
\beq
\overline{x^2} = 2 D t.
\label{12}
\eeq
a result also obtained by Smoluchovski
\cite{smoluchowski1906}.
Indeed, after $\ell$ time steps, the position of the
particle is $x=\ell \zeta$. Therefore, the variance
of $x$ is
$\overline{x^2} = \ell \overline{\zeta^2} = 2 D \ell \tau$.
But $\ell\tau=t$ is the elapsed time and we obtain
(\ref{12}). 

The result that the variance $\overline{\zeta^2}$ 
vanishes with $\tau$ allows to define the random
variable $\xi=\zeta/\sqrt\tau$
which has a finite variance because
$\overline{\xi^2}= \overline{\zeta^2}/\tau=2D$, 
which is finite.
Therefore, defining $\Delta x= x'-x$, then
we may write
\beq
\Delta x = \sqrt{\tau} \xi,
\label{14}
\eeq
an equation that can be understood as the discrete form
in time of the stochastic equation of motion introduced 
by Langevin \cite{langevin1908} and used by 
de Haas-Lorentz \cite{haaslorentz1912} in the context
of the Brownian motion

Let us suppose that the straight line $x$
is discretized in equal spaces $\varepsilon$
and let $p$ be the probability that during
the time interval $\tau$ the particle jumps
to the right and the same probability to the
left. Again the average of $\zeta$ vanishes and
its variance is given by
\beq
\overline{\zeta^2} = 2\varepsilon^2 p.
\eeq
Taking into account that $\overline{\zeta^2} = 2D\tau$,
then, it follows that $p=D\tau/\varepsilon^2$. Therefore,
the transition rate $w=p/\tau$ of jumping to
$\zeta=\varepsilon$ is
\beq
w = \frac{D}{\varepsilon^2}
= \frac{\overline{\xi^2}}{2\varepsilon^2},
\label{20}
\eeq
and the same value for jumping to $\zeta=-\varepsilon$.
We see that the rate  $w$ vanishes as $1/\varepsilon^{2}$
when $\varepsilon$ decreases without limits.

Now we write the master equation associate to the jumping
process which is
\beq
\frac{d \rho(x)}{dt} = w [\rho(x+\varepsilon) - \rho(x)]
+ w[\rho(x-\varepsilon) - \rho(x)],
\eeq
or 
\beq
\frac{d \rho(x)}{dt} = \frac{2D}{\varepsilon^2}
[\rho(x+\varepsilon) - 2\rho(x) + [\rho(x-\varepsilon)].
\eeq
After taking the limit $\varepsilon\to0$, we obtain
the diffusion equation (\ref{10}).

These two lines of reasoning suggest two methods of deriving
the Fokker-Planck equation by a limiting process. The first
considers the discretization of time and is based on the
equation (\ref{14}). The evolution equation of the probability
distribution is obtained by taking the continuous time limit.
The second line of reasoning considers the discretization
of the space of states and is based on the behavior of the
transition rates with the spacing $\varepsilon$ and on the
master equation. The evolution equation of the probability
distribution is obtained by taking the continuous space limit.
Since the second method takes into account the transition
rate it is useful for determining the entropy production
rates, which is defined in terms of the transition rates. 
Our purpose here is to use the second method to
derive the Fokker-Planck from the master equation
and its associated production of entropy from the
expression for entropy production advanced by
Bergmann and Lebowitz \cite{bergmann1955}
and by Schnakenberg \cite{schnakenberg1976}.

\section{Transition rates}

The reasoning that we used above which leaded us to
the form (\ref{20}) for the transition rates are valid
for the symmetric case in which the average of the
random variable $\zeta$ vanishes. Let us consider again
the stochastic motion of a particle along a straight line
and denote by $x$ and $x'$ the position of the particles
at time $t$ and $t+\tau$ respectively. The difference
$\Delta x=x'-x$ is given by
\beq
\Delta x = \zeta,
\eeq
where $\zeta$ is a random variable. If the motion
is not symmetric, then the average of $\zeta$ does
not vanish, which means that
\beq
\overline{\Delta x} = \overline{\zeta}
\eeq
does not vanish.

Taking into account that the ration $\overline{\Delta x}/\tau$
should be finite and that the variance of $\zeta$
should be proportional to $\tau$, as we have argued above,
we may assume the following form for $\zeta$,
\beq
\zeta  = \tau \eta + \sqrt\tau \xi,
\eeq
where $\eta$ is a random variable with finite
mean and $\xi$ is a random variable with zero mean,
and finite variance,
that is,
\beq
\overline{\eta} = f, \qquad \overline{\xi}=0,
\qquad \overline{\xi^2} = g.
\eeq

Next we suppose a discrete motion in which the
particle jumps during a time interval $\tau$
a distance $\varepsilon$ to the right
with probability $p$ and the same distance to the left
with probability $q$. 
The average of the increment $\Delta x=\zeta$  
during the time interval $\tau$ is then
\beq
(p-q)\varepsilon,
\eeq
and its variance is 
\beq
(p+q)\varepsilon^2 - (p-q)^2\varepsilon^2.
\label{13}
\eeq
But the average of $\Delta x$ is the average of $\zeta$
which is $\tau f$ so that
\beq
(p-q)\varepsilon = \tau f,
\eeq
and the variance of $\Delta x$ is the variance of 
$\zeta$ so that
\beq
(p+q)\varepsilon^2 = \tau g,
\eeq
where we have neglected the second term of (\ref{13}) 
which is of the order $\tau^2$. 
Considering that the transition rates $a$ and $b$ for jumping
to the right and to the left, respectively, are given by
$a=p/\tau$ and $b=q/\tau$, we obtain
\beq
(a-b)\varepsilon = f,
\qquad
(a+b)\varepsilon^2 = \tau g,
\eeq
from which follows that the transition rates behaves 
with the spacing $\varepsilon$ as
\beq
a = \frac{g}{2\varepsilon^2} + \frac{f}{2\varepsilon} ,
\qquad
b = \frac{g}{2\varepsilon^2} - \frac{f}{2\varepsilon}.
\label{21}
\eeq
and we remark that $b$ should be strictly positive
to guarantee that $a$ and $b$ be positive.

If we extend the definition of $D$ given by (\ref{11})
to the case where $\overline{\zeta}$ is nonzero then
we should replace $\overline{\zeta^2}$ in this equation
by the variance of $\zeta$, from which we conclude that
\beq
g = 2 D,
\label{11a}
\eeq
that is $g$ is twice the diffusion coefficient.

We remark that the form (\ref{21}) was suggested by 
Gardiner \cite{gardiner1983} as a possible method for
deriving the Fokker-Planck equation from the
master equation, and adds that the method fails
if $g$ vanishes. However, this problem can be 
overcome, as we shall see below.
Concerning not only the derivation of the Fokker-Palnck
but also the associate entropy prodution,  
the form (\ref{21}) was used before
\cite{tome2010,tome2015}. 
Here, we present a more detailed, comprehensive,
and general analysis of the approach based on
(\ref{21}).
%Here, we extend these results to more general situations.

\section{Method}

\subsection{Diagonal case}

We consider a continuous space of states which is 
a vector space $x$ of a given dimension. The components
of the vector $x$ are denoted by $x_i$, 
\beq
x = \sum_i x_i e_i,
\eeq
where $e_i$ is the unit vector in the direction
of the axis $x_i$, that is, $e_i$ has all components
equal to zero except the $i$-th component which is
equal to one.
This space is discretized in cubic cells of
side $\varepsilon$ in such a way that
the possible values taken by $x_i$ are
$\varepsilon n_i$ where $n_i$ is integer.
It is convenient to define the vector $n$ by
\beq
n = \sum_i n_i e_i,
\eeq
so that $x=\varepsilon n$.
We suppose that the stochastic motion on this discrete
space are such that the possible transitions are of the
type $x\to x^i=x+\varepsilon e_i$
and $x\to x^{i-}=x-\varepsilon e_i$.

We denote by $a_i$ and $b_i$ the rates of the transitions
$x\to x^i$ and $x\to x^{i-}$. In accordance with (\ref{21}),
we write
\beq
a_i = \frac{g_i}{2\varepsilon^2} + \frac{f_i}{2\varepsilon},
\qquad
b_i = \frac{g_i}{2\varepsilon^2} - \frac{f_i}{2\varepsilon},
\label{22}
\eeq
where $g_i$ and $f_i$ may depend on $x$ and $g_i$ is
strictly positive, $g_i>0$.
This restriction guarantees that $a_i$ and $b_i$ be
positive for sufficient small $\varepsilon$.

The master equation which gives the evolution of the
probability distribution $\rho$ is given by
\[
\frac{d\rho(x)}{dt} = \sum_i [b_i(x^i)\rho(x^i) - a_i(x)\rho(x)]
\]
\beq
+ \sum_i [a_i(x^{i-})\rho(x^{i-}) - b_i(x)\rho(x)].
\label{24}
\eeq
Replacing (\ref{22}) in this equation, we find
\[
\frac{d\rho(x)}{dt} = -\sum_i \frac1{2\varepsilon}
[f_i(x^i)\rho(x^i) - f_i(x^{i-})\rho(x^{i-})]
\]
\beq
+\frac1{2\varepsilon^2}\sum_i [
g_i(x^i)\rho(x^i) - 2g_i(x)\rho(x) + g_i(x^{i-})\rho(x^{i-})].
\eeq
Taking the limit $\varepsilon\to0$, we obtain
\beq
\frac{\partial\rho}{\partial t}
= - \sum_i \frac{\partial f_i\rho}{\partial x_i}
+ \frac12 \sum_i \frac{\partial^2 g_i\rho}
{\partial x_i^2},
\label{30}
\eeq 
which is the Fokker-Planck equation in
several variables.

\subsection{Production and flux of entropy}

The entropy $S$ of the system described by the master
equation (\ref{24}) is given by the Gibbs formula
in discrete form
\beq
S = - \kappa \sum_n \rho(x) \ln\rho(x),
\eeq
where $\kappa$ is the Boltzmann constant and the
summation in $n$ is the sum in all variables $n_i$,
and $x=\varepsilon n$. Using the master equation,
its time derivative is given by
\beq
\frac{dS}{dt} = \kappa
\sum_{i,n} [b_i(x^i)\rho(x^i) - a_i(x)\rho(x)]
\ln \frac{\rho(x^i)}{\rho(x)}.
\label{27}
\eeq
The entropy is not a conserved quantity so that we may
write its time derivative as
\beq
\frac{dS}{dt} = {\cal P} - \Psi,
\label{26}
\eeq
where ${\cal P}$ is the rate of entropy production and
$\Psi$ is the flux of entropy from the system to the
outside. The production of entropy ${\cal P}$ 
must have the property 
\beq
{\cal P}\geq 0,
\label{25}
\eeq
which is a statement of the second law of thermodynamics.
The splitting of $dS/dt$ into two quantities is
not unique even if we require (\ref{25}). Here we use the
formula advanced by Bergmann and Lebowitz \cite{bergmann1955}
and by Schnakenberg \cite{schnakenberg1976}
which for the present case is given by
\beq
{\cal P} = \kappa
\sum_{i,n} [b_i(x^i)\rho(x^i) - a_i(x)\rho(x)]
\ln \frac{b_i(x^i)\rho(x^i)}{a_i(x)\rho(x)},
\label{36}
\eeq
and it is clear that the property (\ref{25}) holds
because each term of the summation is of the type
$(\nu-\mu)\ln(\nu/\mu)>0$.

If we are given ${\cal P}$, then $\Psi$ can be obtained
from (\ref{26}) and (\ref{27}), and it is
\beq
\Psi = \kappa
\sum_{i,n} [b_i(x^i)\rho(x^i) - a_i(x)\rho(x)]
\ln \frac{b_i(x^i)}{a_i(x)}.
\eeq

Let us define the following quantities
\beq
J_i = - \varepsilon[b_i(x^i)\rho(x^i) - a_i(x)\rho(x)],
\label{34a}
\eeq
\beq
K_i = - \frac1\varepsilon \ln \frac{b_i(x^i)}{a_i(x)},
\label{34b}
\eeq
\beq
L_i = - \frac1\varepsilon \ln \frac{\rho(x^i)}{\rho(x)}.
\label{34c}
\eeq
Then we may write
\beq
\Psi = \kappa\sum_i \int J_i K_i dx,
\label{44a}
\eeq
\beq
{\cal P} = \kappa\sum_i \int J_i (K_i + L_i) dx.
\eeq
\beq
\frac{dS}{dt} = \kappa
\sum_i  \int J_i L_i dx,
\label{44b}
\eeq

Taking the limit $\varepsilon\to 0$, we obtain
\beq
J_i = - \frac12\frac{\partial g_i\rho}{\partial x_i}
+ f_i\rho,
\eeq
\beq
K_i = - \frac1{g_i}\frac{\partial g_i}{\partial x_i}
+ \frac2{g_i} f_i,
\label{44c}
\eeq
\beq
L_i = - \frac1\rho \frac{\partial\rho}{\partial x_i},
\label{44d}
\eeq
and we observe that, 
\beq
K_i + L_i = \frac{2}{g_i\rho} J_i,
\label{29}
\eeq
and the entropy production can be written as
\cite{tome2006,tome2010,vandenbroek2010,spinney2012,tome2015}
\beq
{\cal P} = 2\kappa\sum_i \int \frac{J_i^2}{g_i\rho} dx,
\label{50a}
\eeq
which is the desired result for the entropy production
of a system described by the Fokker-Planck equation
(\ref{30}).

Replacing (\ref{44c}) in (\ref{44a}), we obtain
\beq
\Psi = 2\kappa\sum_i \int \frac{J_i}{g_i}
(f_i - \frac12\frac{\partial g_i}{\partial x_i}) dx,
\label{50b}
\eeq
and replacing (\ref{44d}) in (\ref{44b}), we obtain
\beq
\frac{dS}{dt} = - \kappa
\sum_i  \int \frac{J_i}\rho \frac{\partial\rho}{\partial x_i} dx.
\label{23}
\eeq
This expression can also be obtained directly from the
Fokker-Planck equation (\ref{30}) by writing it
in terms of $J_i$,
\beq
\frac{\partial\rho}{\partial t} = - \sum_i \frac{\partial J_i}
{\partial x_i}.
\label{62}
\eeq
Using the expression for the entropy given by the 
Gibbs formula
\beq
S = - \kappa \int \rho\ln\rho \,dx.
\eeq
we find
\beq
\frac{dS}{dt} = \kappa \sum_i
\int \frac{\partial J_i}{\partial x_i}\ln\rho dx.
\eeq
Integrating by parts and assuming that the integrated 
part vanish, the resulting expression coincides with
(\ref{23}).

\subsection{Flux of heat}

The approach we are presenting is general and describes
stochastic motion of any nature. However, our primary
interest is in describing thermodynamic system in which
case we must establish a connection between the quantities
already introduced and those describing
a thermodynamic system, the main one of which is the
energy function. 
This connection is provided by the Clausius relation
between entropy flux and heat flux. Such a relation
cannot be derived from the results obtained thus far.
Therefore, we need to postulate this relation
or another one which leads directly to it. 

We introduce an energy function $H(x)$ and determine
the time derivative of its average. Using (\ref{62})
we obtain
\beq
\frac{d\la H\ra}{dt} = - \sum_i \int \frac{\partial J_i}
{\partial x_i}H dx
\eeq
After an integration by parts and assuming that the
integrated part vanishes, we find
\beq
\frac{d\la H\ra}{dt} =  \sum_i \int J_i \frac{\partial H}
{\partial x_i}dx
\eeq
Each term of the summation is understood as the
heat flux $\Phi_i$ associated to the process $i$,
\beq
\Phi_i = \int J_i \frac{\partial H}{\partial x_i}dx
\eeq
and 
\beq
\frac{d\la H\ra}{dt} = \Phi = \sum_i \Phi_i.
\eeq

From (\ref{44a}) the entropy flux $\Psi_i$ associated to
the process $i$ is
\beq
\Psi_i = \kappa \int J_i K_i dx,
\eeq
and the total flux $\Psi$ is
\beq
\Psi = \sum_i \Psi_i.
\eeq
The Clausius relation is written as
\beq
\Psi_i = - \frac1{T_i}\Phi_i,
\eeq
where $T_i$ is a positive constant. If the transitions
$x\to x^i$ and its reverse are interpreted as those
induced by the contact of the system with a heat
reservoir, then $T_i$ may be interpreted as the
temperature of the reservoir.
%Within this
%interpretation, the system is understood as being
%in contact with several reservoirs.
This relation is fulfilled if we postulate that
\beq
K_i  = - \beta_i \frac{\partial H}{\partial x_i}
\label{56}
\eeq
where $\beta_i = 1/\kappa T_i$,
which is a relation between the transition rates
and the gradient of $H$.

When all constants $T_i$ are equal it follows that
\beq
\Psi = -\frac1T\Phi
\eeq

In the stationary state, $d\la H\ra/dt=0$ so
that $\Phi=0$ and $\Psi=0$. 
Taking into account that $dS/dt$ also vanishes,
we conclude that the entropy production ${\cal P}$ vanishes.
This means that in fact the stationary state is a state
of thermodynamic equilibrium.
The vanishing ${\cal P}$ implies that each term $J_i$
vanishes, and from (\ref{29}) we see that $K_i+L_i=0$.
Using the expression (\ref{56}) for $K_i$, with $\beta_i=\beta$,
and the expression (\ref{44d}) for $L_i$ we reach
the relation
\beq
\beta \frac{\partial H}{\partial x_i}
+ \frac1\rho \frac{\partial\rho}{\partial x_i} = 0
\eeq
which can be integrated to obtain
\beq
\rho = \frac1Z e^{-H/\kappa T}
\label{46}
\eeq
which is the Gibbs canonical distribution. 

\subsection{General case}

The Fokker-Planck equation (\ref{30}) that we derived does
not contain the nondiagonal terms. This happened because
we considered only transitions along the directions of
the axis. Here we consider a more general case in which
the transitions can be in any direction. The directions
that we consider are those defined by the vectors
\beq
u_k = \sum_i c_{ik} e_i,
\label{35}
\eeq
where the coefficients $c_{ik}$ may depend on $x$. We suppose that
at least two coefficients $c_{ik}$ are nonzero because
the case where just one is nonzero was treated above.

The transitions that we consider are defined by
$x\to y^k = x+\varepsilon u_k$ and 
$x\to y^{k-} = x-\varepsilon u_k$ with rates given
respectively by
\beq
a_k = \frac{g_k}{2\varepsilon^2} + \frac{h_k}{2\varepsilon},
\qquad
b_k = \frac{g_k}{2\varepsilon^2} - \frac{h_k}{2\varepsilon},
\label{22a}
\eeq
where $h_k$ and $g_k$ may depend on $x$.
Again $g_k$ is strictly positive, $g_k>0$.
The master equation is given by
\[
\frac{d\rho(x)}{dt}
= \sum_k [b_k(y^k)\rho_{k}(y^k) - a_k(x)\rho(x)]
\]
\beq
+ \sum_k [a_k(y^{k-})\rho_{k-}(y^{k-}) - b_k(x)\rho(x)].
\label{24a}
\eeq
and we remark that each variable
$x_i$ takes the discretized values given by 
$x_i = \varepsilon n_i$.
We also make the important remark that $\rho_k$, the
probability distribution
of the variable $y^k$, has a functional form 
distinct from the probability distribution $\rho$
of the variable $x$. The same is true for the
distribution $\rho_{k-}$.

Instead of dealing with the master equation we use 
the equation that gives the time derivative of the
average $\la F\ra$ of an arbitrary state function
$F(x)$. This equation is
\[
\frac{d\la F\ra}{dt}
= \sum_{k,n} b_k(x)\rho(x)[F(y^{k-}) - F(x)]
\]
\beq
+ \sum_{k,n} a_k(x)\rho(x)[F(y^k) - F(x)].
\eeq
It can be written in the form
\beq
\frac{d\la F\ra}{dt}
= - \sum_{k,n} [b_k(y^k)\rho_k(y^k) - a_k(x)\rho(x)][F(y^k) - F(x)].
\eeq
Let us define
\beq
J_k(x) = - \varepsilon[b_k(y^k)\rho_k(y^k) - a_k(x)\rho(x)],
\label{39}
\eeq
and 
\beq
A_k(x) = \frac1{\varepsilon}[F(y^k) - F(x)],
\label{39d}
\eeq
then we may write
\beq
\frac{d\la F\ra}{dt} = \sum_{k,n} J_k(x) A(x).
\label{40}
\eeq

Taking the limit $\varepsilon\to0$, the expression
(\ref{39d}) becomes
\beq
A_k = \sum_i \frac{\partial F}{\partial x_i}c_{ik},
\eeq
As to the quantity $J_k$ it is finite in this limit
as we will se below and
\beq
\frac{d\la F\ra}{dt} = \sum_k \int J_k(x) A(x)dx.
\eeq
After an integration by parts and assuming that the
integrated part vanishes we obtain
\beq
\frac{d\la F\ra}{dt} 
= - \sum_{ik} \int F \frac{\partial c_{ik} J_k}{\partial x_i} dx.
\eeq
Taking into account the $F$ is arbitrary, we find
\beq
\frac{\partial\rho}{\partial t} 
= - \sum_{ik} \frac{\partial c_{ik} J_k}{\partial x_i}.
\label{40a}
\eeq

Now we have to determine the expression for $J_k$
in the limit $\varepsilon\to0$. To this end, 
we start by replacing (\ref{22a}) in (\ref{39}),
\beq
J_k(x) = h_k(x)\rho(x)
- \frac1{2\varepsilon}[g_k(y^k)\rho_k(y^k) - g_k(x)\rho(x)],
\eeq
where we have neglected the terms of order $\varepsilon$.
To proceed further,
it is necessary to take into accont the following observation.
The distributions $\rho(x)$ and $\rho_{k}(y^k)$ of the variables
$x$ and $y^k$ are distinct in general but they become identical in 
the limit $\varepsilon\to0$. This allows us to write
$\rho_{k}(y^k)=\rho(y^k)[1+\varepsilon w_k(x)]$,
up to order $\varepsilon$, where $w_k(x)$ will be determined
shortly. The quantity $J_k$ then becomes
\[
J_k(x) = h_k(x)\rho(x) - \frac12 g_k(x)\rho(x) w_k(x)
\]
\beq
- \frac1{2\varepsilon}[g_k(y^k)\rho(y^k) - g_k(x)\rho(x)]
\eeq
where we have neglected the terms of order $\varepsilon$.
Taking the limit $\varepsilon\to0$ of this expression we obtain
\beq
J_k = (h_k - \frac12 g_k w_k)\rho
-\frac12 \sum_i \frac{\partial g_k\rho}{\partial x_i}c_{ik},
\label{42}
\eeq
and it remains to determine the quantity $w_k$.

The distributions $\rho(x)$ and $\rho_{k}(y^k)$
are related by $\rho_{k}(y^k){\cal J}_k(x)=\rho(x)$
where ${\cal J}_k$ is the Jacobian of the transformation
$x\to y_k=x+\varepsilon u+k$. Up to order $\varepsilon$,
it is given by
\beq
{\cal J}_k = 1 + \varepsilon \sum_i \frac{\partial c_{ik}}{\partial x_i} 
\eeq
which allows us to conclude that $w_k$ is
\beq
w_k = \sum_i \frac{\partial c_{ik}}{\partial x_i} 
\eeq
Replacing this result in (\ref{42}), we obtain
\beq
J_k = h_k \rho
-\frac12 \sum_i \frac{\partial g_k c_{ik} \rho}{\partial x_i},
\label{47}
\eeq

The substitution of this expression in (\ref{40a}
gives
\beq
\frac{\partial\rho}{\partial t} 
= - \sum_{ik} \frac{\partial c_{ik}  h_k \rho }{\partial x_i}
+ \frac12 \sum_{ijk} \frac{\partial }{\partial x_i}
c_{ik} \frac{\partial g_k c_{jk} \rho}{\partial x_j}.
\eeq
Defining the functions $f_i$ and $g_{ij}$ by
\beq
f_i = \sum_k (c_{ik}h_k + \frac12\sum_j
g_k \frac{\partial c_{ik}}{\partial x_j}),
\eeq
and
\beq
g_k = \sum_k c_{ik}g_k c_{jk},
\eeq
then it can be written in the simplified form
\beq
\frac{\partial\rho}{\partial t} 
= - \sum_i \frac{\partial f_i \rho }{\partial x_i}
+ \frac12 \sum_{ij} \frac{\partial^2 g_{ij} \rho}{\partial x_i\partial x_j},
\eeq
which is a Fokker-Planck equation.

\subsection{Flux and production of entropy}

As before, the entropy production is given by
\beq
{\cal P} = \kappa\sum_k \int J_k M_k dx,
\eeq
where 
\beq
M_k = - \frac1\varepsilon \ln \frac{b_k(y^k)\rho_k(y^k)}{a_k(x)\rho(x)}
\eeq
To establish the formula for flux of entropy $\Psi$,
we use the formula $dS/dt={\cal P}-\Psi$. Multiplying
(\ref{24a}) by $-\kappa \ln\rho(x)$ and summing in $n$,
we obtain
\beq
\frac{dS}{dt} = \kappa \sum_k \int J_k L_k
\eeq
where $J_k$ is the quantity (\ref{39}), and $L_k$ is
\beq
L_k = -\frac1\varepsilon \ln \frac{\rho(y^k)}{\rho(x)}
\eeq
The flux of entropy is thus 
\beq
\Psi = \kappa\sum_k \int J_k K_k dx,
\label{54e}
\eeq
where $K_k=M_k-L_k$ and given by
\beq
K_k = - \frac1\varepsilon \ln \frac{b_k(y^k)}{a_k(x)}
- \frac1\varepsilon \ln \frac{\rho_k(y^k)}{\rho(y^k)}
\eeq
which can be written as
\beq
K_k = - \frac1\varepsilon \ln \frac{b_k(y^k)}{a_k(x)}
- w_k(x)
\eeq

The limit $\varepsilon\to0$ of the first term o $K_k$
is obtained  by first using (\ref{22a}) to write
\beq
K_k = - \frac1\varepsilon \ln \frac{g_k(y^k)- h_k(y^k)\varepsilon }
{g_k(x) + h_k(x) \varepsilon} - w_k(x)
\eeq
After taking the limit we get
\beq
K_k =  \frac{2 h_k}{g_k}
- \frac{1}{g_k}\sum_i \frac{\partial g_k}{\partial x_i}c_{ik}
- \sum_i \frac{\partial c_{ik}}{\partial x_i} 
\eeq
or
\beq
K_k =  \frac{2 h_k}{g_k}
- \frac{1}{g_k}\sum_i \frac{\partial g_k c_{ik}}{\partial x_i}
\label{42a}
\eeq

The limit of $L_k$ is
\beq
L_k = - \frac1{\rho}\sum_i \frac{\partial \rho}{\partial x_i}c_{ik},
\label{42b}
\eeq 
and we see that
\beq
K_k + L_k =  \frac{2 J_k}{g_k\rho}
\eeq
Replacing this result in the expression for the production of
entropy we obtain 
\beq
{\cal P} = 2\kappa\sum_k \int \frac{J_k^2}{g_k\rho} dx,
\eeq
and we see that it is nonnegative since $g_k\geq0$,
as expected.
Replacing $K_k$ in the expression for the flux of entropy
we find
\beq
\Psi = \kappa\sum_k \int \frac{J_k}{g_k} (2 h_k
- \sum_i \frac{\partial g_k c_{ik}}{\partial x_i}) dx,
\eeq

It is worth mentioning that in the present case
the relation (\ref{56}) becomes
\beq
K_k = - \beta_k \sum_i
\frac{\partial H}{\partial x_i} c_{ik}.
\label{56a}
\eeq
$\beta_k=1/\kappa T_k$,
which leads us to the Clausius relation
\beq
\Psi_k = -\frac1{T_k}\Phi_k,
\eeq 
between the entropy flux 
\beq
\Psi_k = \kappa\int J_k K_k dx,
\eeq
an the heat flux 
\beq
\Phi_k =  \kappa \sum_i \int
\frac{\partial H}{\partial x_i} c_{ik} dx.
\eeq
associated to process $k$.

If all $T_k$ are equal we may use the same reasoning
employed before to show that in the stationary state
the system will be in equilibrium with the Gibbs
canonical distribution (\ref{46}).

\subsection{Change of two variables}

Here we restrict the possible directions to those that
lay on planes of the space of states. 
We use a more appropriate notation such that the index
$k$ of the vector $u_k$ is replaced by the double index
$kl$, meaning that the direction lays in the plane 
$(x_k,x_l)$. Accordingly,
\beq
u_{kl} = c_{kl} e_k + c_{lk} e_l,
\eeq
and write $h_{kl}$ and $g_{kl}$ in the place of
$h_k$ and $g_k$.
%tirei o chapeu de g_{kl}

The formulas (\ref{47}), (\ref{42a}), and (\ref{42b}) become
\beq
J_{kl} = h_{kl}\rho 
- \frac12(\frac{\partial g_{kl}c_{kl} \rho}{\partial x_k}
+ \frac{\partial g_{kl} c_{lk}\rho}{\partial x_l} ),
\eeq
\beq
K_{kl} = \frac2{g_{kl}} h_{kl}
- \frac1{g_{kl}}(\frac{\partial g_{kl}c_{kl}}
{\partial x_k} 
+ \frac{\partial g_{kl}c_{lk}}{\partial x_l}),
\eeq
\beq
L_{kl} = - \frac1\rho (\frac{\partial\rho}{\partial x_k}
c_{kl} + \frac{\partial\rho}{\partial x_l} c_{lk}).
 \eeq

The formulas for ${\cal P}$, $\Psi$, and $dS/dt$
are given by
\beq
{\cal P} = \kappa \sum_{k(\neq)l} \int \frac{J_{kl}^2}
{g_{kl}\rho} dx,
\eeq
\beq
\Psi = \frac\kappa2 \sum_{k(\neq)l} \int J_{kl} K_{kl} dx,
\label{55b}
\eeq
\beq
\frac{dS}{dt} = \frac\kappa2 \sum_{k(\neq)l} 
\int J_{kl} L_{kl} dx,
\eeq
and the equation (\ref{40a}) for the evolution of $\rho$ is
\beq
\frac{\partial\rho}{\partial t} 
= - \sum_{k\neq l}\frac{\partial c_{kl} J_{kl}}{\partial x_k},
\eeq
which can be written as
\[
\frac{\partial\rho}{\partial t} 
= - \sum_{k\neq l}\frac{\partial }{\partial x_k}c_{kl} h_{kl}\rho
\]
\beq
+\frac12 \sum_{k\neq l}\frac{\partial }{\partial x_k}
(c_{kl}\frac{\partial g_{kl}c_{kl} \rho}{\partial x_k}
+ c_{kl}\frac{\partial g_{kl} c_{lk}\rho}{\partial x_l} )
\eeq

\subsection{Example with zero entropy flux}

An example of a system with zero entropy flux
is the one in which $g_{kl}=g$ is a
constant, $h_{kl}=0$ and $c_{kl}$ fulfills the
condition
\beq
\frac{\partial c_{kl}} {\partial x_k} 
+ \frac{\partial c_{lk}}{\partial x_l} = 0.
\label{60}
\eeq
Using this condition $J_{kl}$ becomes
\beq
J_{kl} =  
- \frac{g}2(\frac{\partial \rho}{\partial x_k} c_{kl}
+ \frac{\partial\rho}{\partial x_l} c_{lk}),
\eeq
and $L_{kl} = 2J_{jk}/g\rho$.
Replacing $J_{kl}$ in the expression for the production
of entropy we reach the result
\beq
{\cal P} = \frac{\kappa g}{4}\sum_{k(\neq)l} \int \frac1{\rho}
(\frac{\partial \rho}{\partial x_k} c_{kl}
+ \frac{\partial\rho}{\partial x_l} c_{lk})^2 dx.
\eeq
Since $\Psi=0$, then
\beq
\frac{dS}{dt} = {\cal P}.
\eeq

An interesting case is the one in which the coefficients
$c_{kl}$ are
\beq
c_{kl} = \alpha_{kl} x_l, \qquad \alpha_{kl}=-\alpha_{lk}.
\eeq
which satisfies the condition (\ref{60}).

The expression for $J_{kl}$ becomes
\beq
J_{kl} =  
- \frac{g}2 \alpha_{kl} (x_l \frac{\partial \rho}{\partial x_k}
- x_k \frac{\partial\rho}{\partial x_l}),
\eeq
and the production of entropy is
\beq
{\cal P} = \frac{\kappa}{2}\sum_{k(\neq)l}\gamma_{kl}
\int \frac1{\rho}
(x_l\frac{\partial \rho}{\partial x_k}
- x_k\frac{\partial\rho}{\partial x_l})^2 dx,
\label{63}
\eeq
and the equation of evolution of $\rho$ is
\beq
\frac{\partial\rho}{\partial t}
= \sum_{k(\neq)l} \gamma_{kl}
\frac{\partial}{\partial x_k}\left(
x_l^2 \frac{\partial\rho}{\partial x_k}
-  x_l x_k\frac{\partial\rho}{\partial x_l} 
\right).
\label{58}
\eeq
where $\gamma_{kl}=g \alpha_{kl}^2/2$.

\section{Unidirectional transitions}

The transition rates $a_k$ and $b_k$ given by equation
(\ref{22a}) are restricted to the case where $g_k$,
which is related to the variance of the independent
random variable, is strictly positive. The case where  
$g_k$ is zero would yield either $a_k$ or $b_k$.
Therefore, the case $b_k$ equal to zero is possible only
if either $a_k$ is equal to zero or $b_k$ is equal to zero.
This case corresponds
to transitions that do not have their reverses,
which we call unidirectional transitions.

We suppose that a transition occurs along a certain
direction given by the vector $u_k$ defined by
(\ref{35}). The rate of the transition
$x\to y^k=x+\varepsilon u_k$ is now given by
\beq
a_k = \frac{h_k}{\varepsilon},
\label{38}
\eeq
where $h_k\geq 0$, and the rate of the reverse vanishes.

The master equation is given by
\beq
\frac{d\rho(x)}{dt} = \sum_k [a_k(y^{k-})\rho_{k-}(y^{k-})
- a_k(x)\rho(x)],
\label{37}
\eeq
and we recall that $y^{k-}=x-\varepsilon u_k$.
As before, we consider the time derivative of 
the average of an arbitrary function $F(x)$, which is
\beq
\frac{d\la F\ra}{dt} = \sum_k \sum_n a_k(x)\rho(x)
[F(y^k) - F(x)].
\eeq
Replacing the transition rate in this equation,
and taking the limit, we obtain
\beq
\frac{d\la F\ra}{dt} = \sum_{ik} \int h_k\rho
\frac{\partial F}{\partial x_i}c_{ik} dx.
\eeq
After an integration by parts and assuming that the
integrated part vanishes, we obtain
\beq
\frac{d\la F\ra}{dt} = - \sum_{ik} \int 
\frac{\partial h_k c_{ik}\rho }{\partial x_i} F dx.
\eeq
Taking into account that $F$ is arbitrary, we find
\beq
\frac{\partial\rho}{\partial t}
= - \sum_{ik} \frac{\partial h_k c_{ik}\rho}{\partial x_i}.
\label{52}
\eeq
It is convenient to write this equation as
\beq
\frac{\partial\rho}{\partial t}
= - \sum_i \frac{\partial f_i^{u}\rho}{\partial x_i},
\label{31}
\eeq
where
\beq
f_i^u = \sum_k c_{ik} h_k.
\eeq

From the equation (\ref{52}), the time derivative
of the entropy
\beq
S = - \kappa \int \rho \ln\rho dx
\eeq
is given by
\beq
\frac{dS}{dt} = - \kappa 
\sum_i \int f_i^u\frac{\partial\rho}{\partial x_i} dx.
\eeq
After an integration by parts and assuming that the integrated
part vanishes, we reach the result
\beq
\frac{dS}{dt} = \kappa \sum_i
\la \frac{\partial f_i^u}{\partial x_i}\ra,
\label{49}
\eeq
which is the average of the divergence of the vector field with
components $f_j^u$. Thus if the vector field has zero divergence
then $dS/dt=0$ and the entropy is constant.

\section{Stochastic motion in phase space}

\subsection{Method}

The choice of a space of states depends on the nature
of the system one wishes to describe. In the case of a
mechanical system the space of states could be the space
of the positions of the particles of the system,
known as space of configurations. However, the Newtonian
equations of motion are second order in time which means
that the positions alone cannot define the initial state
and therefore cannot define a space of states. To remedy
this problem it suffices to supplement the positions with
the moments of the particles. The space of states thus
 becomes composed of the space of configurations $x$ and
 the space of momenta $p$, combination known as phase space.
The original equations of motion, which are second order
in time, become a set of equations of first order in time,
the number of which is doubled.

The stochastic motion in phase space is assumed to
consist of two classes of transitions. The first
class corresponds to the unidirectional transitions
and are of two types that are related to the positions
and momenta, respectively. The first is defined by
$x_i\to x_i'=x_i + \sigma_i'\varepsilon'$ where
$\sigma_i'$ is the sign of $v_i$, with rate
\beq
w_i = \frac{|v_i|}{\varepsilon'}, \qquad
v_i = \frac{\partial H}{\partial p_i},
\label{45a}
\eeq 
where $\varepsilon'$ is the spacing of the configuration
space, and $v_i$ is the velocity. The second is the
transition defined by $p_i\to p_i'=p_i + \sigma_i\varepsilon$
where $\sigma_i$ is the sign of $f_i^c$, occurring with rate
\beq
c_i = \frac{|f_i^c|}{\varepsilon}, \qquad
f_i^c = - \frac{\partial H}{\partial x_i},
\label{45b}
\eeq 
where $\varepsilon$ is the spacing of the space of
momenta, and $f_i^c$ is a conservative force.

The second class corresponds to the
transitions $p_i\to p_i'= p_i+\varepsilon$ and
$p_i\to p_i''= p_i-\varepsilon$, with rates
\beq
a_i = \frac{g_i}{2\varepsilon^2} + \frac{f_i}{2\varepsilon},
\qquad
b_i = \frac{g_i}{2\varepsilon^2} - \frac{f_i}{2\varepsilon},
\label{45c}
\eeq
and we recall that $g_i\geq0$.

Instead of deriving the Fokker-Planck equation 
as we did before, we adapt the previous results
to the present case. Accordingly, we use the
results (\ref{52}) and (\ref{30}) to obtain
the equation for the time evolution of $\rho$
\[
\frac{d\rho}{dt} = 
- \sum_i \frac{\partial |v_i|\sigma_i' \rho}{\partial x_i}
- \sum_i \frac{\partial |f_i^c|\sigma_i\rho}{\partial p_i} 
\]
\beq
- \sum_i \frac{\partial f_i \rho}{\partial p_i} 
+ \frac12 \sum_i \frac{\partial^2 g_i \rho}{\partial p_i^2}.
\eeq
Taking into account that $v_i=|v_i|\sigma_i'$ and that
$f_i^c=|f_i^c|\sigma_i$, we may write
\[
\frac{d\rho}{dt} = 
- \sum_i \frac{\partial v_i \rho}{\partial x_i}
- \sum_i \frac{\partial f_i^c\rho}{\partial p_i} 
\]
\beq
- \sum_i \frac{\partial f_i \rho}{\partial p_i} 
+ \frac12 \sum_i \frac{\partial^2 g_i \rho}{\partial p_i^2}.
\label{48b}
\eeq
We observe that
\beq
\sum_i (\frac{\partial v_i \rho}{\partial x_i}
+ \frac{\partial f_i^c\rho}{\partial p_i}) =
\sum_i (v_i \frac{\partial\rho}{\partial x_i}
+ f_i^c \frac{\partial \rho}{\partial p_i}),
\eeq
which is equal to
\beq
\sum_i (\frac{\partial \rho}{\partial x_i}
\frac{\partial H}{\partial p_i}- \frac{\partial \rho}{\partial p_i}
\frac{\partial H}{\partial x_i}) = \{\rho,H\},
\eeq
the Poisson bracket between $\rho$ and $H$, and the equation
(\ref{48b}) simplifies to
\beq
\frac{\partial\rho}{\partial t} = \{H,\rho\}
- \sum_i \frac{\partial f_i\rho}{\partial p_i}
+\frac12\sum_i \frac{\partial^2 g_i\rho}{\partial p_i^2}.
\label{48}
\eeq
This equation can be written as
\beq
\frac{\partial\rho}{\partial t} = \{H,\rho\}
- \sum_i \frac{\partial J_i}{\partial p_i},
\label{48a}
\eeq
where
\beq
J_i = f_i\rho - \frac12\frac{\partial g_i\rho}{\partial p_i}.
\eeq

The time derivative of the entropy
\beq
S = - \kappa \int \rho\ln\rho dxdp
\eeq
is obtained from 
equation (\ref{48a}) and is given by
\beq
\frac{dS}{dt} = 
- \sum_i \int\frac{J_i}{\rho}
\frac{\partial \rho}{\partial p_i} dxdp,
\eeq
result obtained by an integration by parts and
assuming that the integrated part vanishes. The
term associated to the Poisson brackets disappears.

The production of entropy and the flux of entropy
are given by equations (\ref{50a}) and (\ref{50b}),
which for the present case are, respectively,
\beq
{\cal P} = 2\kappa\sum_i \int \frac{J_i^2}{g_i\rho} dx dp,
\label{51a}
\eeq
\beq
\Psi = 2\kappa\sum_i \int\frac{J_i}{g_i}
(f_i - \frac12\frac{\partial g_i}{\partial x_i}) dx dp.
\label{51b}
\eeq

\subsection{Contact with several heat reservoirs}

To describe a system in contact with several heat
reservoir we use the relation (\ref{56}) and
$K_i$ given by (\ref{44c}). For the present case
it is given by
\beq 
\frac2{g_i} f_i
- \frac1{g_i}\frac{\partial g_i}{\partial p_i}
= - \frac1{\kappa T_i} \frac{\partial H}{\partial p_i}.
\label{61}
\eeq
Taking into account that $v_i=\partial H/\partial p_i$,
we write this condition as
\beq 
\frac2{g_i} f_i
- \frac1{g_i}\frac{\partial g_i}{\partial p_i}
= - \frac{v_i}{\kappa T_i}.
\eeq

There are several ways to choose $f_i$ and $g_i$
so as to meet this condition. The usual choice is
\beq
g_i = 2\gamma_i \kappa T_i, \qquad
f_i = -\gamma_i v_i.
\label{59}
\eeq
that is, $g_i$ is a constant proportional to $T_i$,
and $f_i$ proportional to $v_i$.
It is worth mentioning that Kramers \cite{kramers1940}
raised the possibility of other choice such as
$f_i =-\gamma_i v_i -\alpha_i v_i^2$. However, he states
that he did not know if these more complicated cases
might have some physical applications. He further states
that the simplest choice (\ref{59})
corresponds to the Einstein case.

We recall that, according to (\ref{11a}), $g_i$ is
twice the diffusion constant $D_i$. Thus according
to the choice (\ref{59}), 
\beq
D_i = \gamma_i \kappa T_i
\eeq
which is the relation between the diffusion constant
and the temperature introduced by Sutherland
\cite{sutherland1905} and Einstein \cite{einstein1905}.
Thus according to the present approach the
Sutherland-Einstein relation is a consequence of
the postulate (\ref{56}).

If $g_i$ and $f_i$ are given by (\ref{59}), then
\beq
J_i = -\gamma_i v_i \rho -
\gamma_i \kappa T_i\frac{\partial\rho}{\partial p_i},
\eeq
and the equation (\ref{48a})
that gives the tme evolution of $\rho$ is
\beq
\frac{\partial\rho}{\partial t} = \{H,\rho\}
+ \sum_i \gamma_i \frac{\partial v_i \rho}{\partial p_i}
+ \kappa \sum_i \gamma_i T_i
\frac{\partial^2\rho}{\partial p_i^2}
\label{28}
\eeq
For one degree of freedom this
equation was introduced by Kramers \cite{kramers1940}.

The associated production of entropy and the flux
of entorpy associated with (\ref{28}) are obtained
from (\ref{51a}) and (\ref{51b}), respectively,
and are given by \cite{tome2010,tome2015}
\beq
{\cal P} = \sum_i \frac1{\gamma_i T_i}
\int \frac{J_i^2}{\rho} dx dp,
\eeq
and by \cite{tome2010,tome2015}
\beq
\Psi = - \sum_i \frac1{T_i}\int  J_i v_i  dx dp,
\eeq
After an integration by parts and assuming that
the integrated part vanish the flux of entroy
can be written as \cite{tome2010,tome2015}
\beq
\Psi =\sum_i \frac{\gamma_i}{mT_i}
(m\la v_i^2\ra - \kappa T_i ).
\eeq

Let us determine the conditions under which the
system is in a state of thermodynamic equilibrium in the
stationary state. For this to happen, it is necessary that
the entropy production vanishes which means that
$J_i=0$ or
\beq
f_i\rho - \frac12\frac{\partial g_i\rho}{\partial p_i} = 0,
\eeq
which is equivalent to
\beq
\frac{\partial \ln \rho}{\partial p_i}
= \frac2{g_i}(f_i - \frac12\frac{\partial g_i}{\partial p_i}).
\label{53}
\eeq
Using (\ref{61}) we find the condition
\beq
\frac{\partial \ln \rho}{\partial p_i}
= - \frac1{\kappa T_i} \frac{\partial H}{\partial p_i},
\eeq
which is satisfied if all $T_i$ are equal. Setting
the common value as $T$ we obtain after integration
\beq
\rho = \frac1Z e^{-H/\kappa T},
\eeq
which is the Gibbs canonical
distribution. 

\subsection{Isolated systems}

An isolated system does not exchange heat with the
environment and is understood  as a system such that
$H$ is strictly conserved, which means that the flux
of heat vanishes, $\Phi=0$. Usually it is
described by the Liouville equation 
\beq
\frac{\partial\rho}{\partial t} = \{H,\rho\},
\eeq
and as we have seen above,
$dS/dt=0$ so that the entropy is conserved. 

It is possible however to construct a stochastic
dynamics such that $\Phi=0$. To this end, we consider
a stochastic dynamics composed of two types of
transitions. The first type are the unidirectional
transitions defined by (\ref{45a}) and (\ref{45b}).
These transitions lead us to a term to the time evolution
of $\rho$ consisting of the Poisson brackets. 
The second type of transitions are those 
that we have used in the example above.
In this example, any two pairs of variables
could change in a transition. Here we consider
that only pairs of variables that may change
are the pairs composed by momenta.

Using (\ref{58}), the equation that gives the
time evolution of $\rho$ is
\beq
\frac{\partial\rho}{\partial t} = \{H,\rho\}
+ \sum_{i(\neq)j} \gamma_{ij}
\frac{\partial}{\partial p_i}(
p_j^2 \frac{\partial\rho}{\partial p_i}
-  p_j p_i\frac{\partial\rho}{\partial p_j}).
\label{57}
\eeq
and can be written as 
\beq
\frac{\partial\rho}{\partial t} = \{H,\rho\}
- \sum_i \frac{\partial J_i}{\partial p_i},
\label{57a}
\eeq
where 
\beq
J_i = - \sum_{j(\neq i)} \gamma_{ij}p_j
(p_j \frac{\partial\rho}{\partial p_i}
- p_i\frac{\partial\rho}{\partial p_j}).
\eeq

We have to show now that $H$ is conserved.
To this end we write the time derivative of a
state function $F(x,p)$. Using (\ref{57a}), we find
\beq
\frac{d\la F\ra}{dt} = \int \{F,H\}\rho dxdp
+ \sum_i \int J_i \frac{\partial F}{\partial p_i} dxdp,
\eeq
a result obtained by integration by parts and assuming
that the integrated parts vanish. Replacing $J_i$
in this expression and using the same procedure 
of integration, we find
\[
\frac{d\la F\ra}{dt} = \int \{F,H\}\rho dxdp
\]
\beq
+ \sum_{i(\neq)j}  \gamma_{ij} \int \rho
\frac{\partial}{\partial p_i}p_j(p_j
\frac{\partial F}{\partial p_i}
-  p_i\frac{\partial F}{\partial p_j}) dxdp.
\eeq
If a quantity is conserved then it is necessary that the
terms that multiply $\rho$ should vanish. Therefore
for $H$ to be conserved it suffices that 
\beq
p_j \frac{\partial H}{\partial p_i}
-  p_i\frac{\partial H}{\partial p_j} = 0,
\eeq
because the Poisson term vanishes.
This requirement is fulfilled for $H$
given by
\beq
H = \sum_i \frac{p_i^2}{2m} + V(x),
\eeq
which is the energy function for an interacting
system whose kinetic energy are translational,

Is it is worth mentioning that unlike the Liouville,
for which the entropy is strictly constant,
the isolated system described by (\ref{57})
exhibits increasing entropy, and thus in accordance
with the thermodynamics of isolated systems. 
To show that this is indeed the case,
we determine the production of entropy
associated to the second term of (\ref{57})
which is obtained from (\ref{63}) and is
\beq
{\cal P} = \frac{\kappa}{2}\sum_{i(\neq)j}\gamma_{ij}
\int \frac1{\rho}
(p_j\frac{\partial \rho}{\partial p_i}
- p_i\frac{\partial\rho}{\partial p_j})^2 dx dp.
\label{64}
\eeq

\section{Conclusion}

We have derived the Fokker-Planck equation by a method
based on the discretization of the space of states
and using appropriate transition rates to 
construct a master equation. The Fokker-Planck 
equation is then derived by taking the continuum limit.
A transition rate is a sum of two terms.
One of then is proportional to the variance of
the independent stochastic variables and inversely
proportional to the square of the spacing.
The other is proportional to the mean of the
stochastic variable and inversely proportional
to the spacing. 
This method allows the calculation of the production
of entropy from the formula (\ref{36}) and also the
flux of entropy. The continuum limit of these
expression gives the production of entropy and the
flux of entropy associated to the Fokker-Planck
equation.

We have applied the results to the case where the
space of states is the phase space and considered
two particular cases. The contact with a system
with several heat reservoirs, in which case the system
may be found in a state out of equilibrium with
a continuous production of entropy. The other
case corresponds to an isolated system in which
the energy is strictly conserved. In this situation
the equation predicts an increase of entropy
in accordance with the thermodynamics of 
isolated systems.

\end{document}